      [1999/12/01 v1.4c Il Nuovo Cimento]
\documentclass{cimento}

\usepackage{bm}
\usepackage{graphicx}  

\title{Azimuthal distributions of pions inside a jet in hadronic collisions}
\author{U.~D'Alesio\from{ins:uni}\ETC,
F.~Murgia\from{ins:infn}
        \atque
C.~Pisano\from{ins:uni}\thanks{Speaker. Talk given at Third International Workshop on Transverse Polarization Phenomena in Hard Scattering (Transversity 2011), Veli Lo\v{s}inj, Croatia, 29 August~-~2 September~2011.}}
\instlist{\inst{ins:uni} Dipartimento di Fisica, Universit\`a di Cagliari, Cittadella Universitaria, Monserrato, Italy
  \inst{ins:infn}  Istituto Nazionale di Fisica Nucleare, Sezione di Cagliari, C.P.\ 170, Monserrato, Italy}
\PACSes{
\PACSit{13.85.Ni}{Inclusive production with identified hadrons}\PACSit{13.88.+e}{Polarization in interactions and scattering}}
\begin{document}

\maketitle

\begin{abstract} Using a generalized parton model approach including spin and intrinsic parton motion effects, and assuming the validity of factorization for large $p_T$ jet production in hadronic collisions, we study the azimuthal
distribution around the jet axis of leading pions, produced in the jet fragmentation process. We identify the observable leading-twist azimuthal asymmetries, which are generated by all the physically allowed combinations of transverse momentum  dependent (TMD) parton distribution and fragmentation functions. In particular, we show how one can isolate the Collins and Sivers contributions, and suggest a test of the process dependence of the Sivers function by considering the effect of color-gauge invariant initial and final state interactions.  
\end{abstract}

\section{Introduction}
Transverse single-spin and azimuthal asymmetries in high-energy hadronic
reactions have raised a lot of interest in the last years
(see e.g.~Ref.~\cite{D'Alesio:2007jt} and references therein).
In particular, the huge spin asymmetries measured in the inclusive 
forward production of pions in high-energy proton-proton  collisions, at moderately large transverse momentum, cannot be explained in the realm of
leading-twist (LT) perturbative QCD (pQCD),
based on the usual collinear factorization theorems. Out of the theoretical approaches proposed in order to account for these measurements, in the following we will  adopt the so-called transverse momentum dependent (TMD) formalism, which takes into account spin and intrinsic parton motion effects assuming a pQCD factorization scheme. In this framework, single-spin and azimuthal asymmetries are generated by TMD polarized partonic distribution and fragmentation functions, among which the most relevant from a phenomenological point of view are the Sivers distribution~\cite{Sivers:1989cc} and, for transversely polarized quarks, the Boer-Mulders distribution~\cite{Boer:1997nt} and the Collins fragmentation function~\cite{Collins:1992kk} (similar functions can be defined for linearly polarized gluons, see e.g.~Ref.~\cite{Anselmino:2005sh}). 

Azimuthal asymmetries in the distribution of pions inside a large transverse momentum jet are quite interesting observables~\cite{D'Alesio:2010am}
and are presently under active investigation at the Relativistic Heavy Ion Collider (RHIC).
In contrast to inclusive  pion 
production, where several underlying competing mechanisms (mainly
the Sivers and Collins ones) cannot be separated, and  in close 
analogy with the semi-inclusive deep inelastic scattering (SIDIS) case, one could discriminate among different effects by taking suitable moments of these asymmetries.  In principle, quark and gluon originating jets can also be distinguished, at least in some kinematic regimes. Furthermore, information on the size of TMD distributions and fragmentation functions can be gained, in a kinematic region in which they are still poorly known.
This would  be very helpful in clarifying the role played by the quark(gluon) Sivers distribution and by the Collins(-like) fragmentation function in the sizable single spin asymmetries observed at RHIC for single inclusive pion production.
A similar analysis where transverse partonic motion was considered only in the fragmentation process, aimed at a study of the universality of the Collins function for
quarks, was presented in Ref.~\cite{Yuan:2007nd}.
Our approach, named generalized parton model (GPM), has in principle a richer structure in the observable azimuthal asymmetries, because intrinsic motion is taken into account in the initial hadrons as well. Preliminary data from the STAR collaboration at RHIC seem to support our model, since they 
report on a Sivers asymmetry for neutral pions larger than zero
\cite{Poljak:2010tm,Poljak:2011vu} and compatible with our predictions.
However, since factorization has not been proven in this case, but is rather taken as a reasonable phenomenological assumption,
the validity of the scheme and the universality of the TMD
distributions involved require a severe scrutiny
by further comparison with experiments. 

\section{\label{sec-results} Theoretical framework}
We consider the process $p^{\uparrow}p\to{\rm jet}+\pi+X$, with one of the protons in a transverse spin state described by the four-vector $S$. We work in the $p\,p$ c.m. frame, where the polarized proton moves along the  $+\hat{\bm{Z}}_{\rm cm}$ direction, and define $(XZ)_{\rm cm}$ as the
production plane containing the colliding beams and the observed jet, with $(\bm{p}_{\rm j})_{X_{\rm cm}}>0$. In this frame $S = (0, \cos\phi_{S},\sin\phi_{S},0) $ and $p_{\rm j} = 
p_{{\rm j}\,T}(\cosh \eta_{\rm j},1,0,\sinh \eta_{\rm j})$,
where $\eta_{\rm j} = -\log[\tan(\theta_{\rm j}/2)]$ is the jet (pseudo)rapidity. We denote by $z$ and $\bm{k}_{\perp\pi}$ respectively the light-cone momentum fraction and the transverse momentum of the observed pion inside the jet w.r.t.\  the jet (fragmenting parton) direction of motion. 
Calculations have been performed within the GPM framework at leading-order (LO) in pQCD utilizing the helicity formalism.
More details can be found in Ref.~\cite{D'Alesio:2010am}.

The final expression for the single-transverse polarized cross 
section has the following general structure:
\begin{eqnarray}
2{\rm d}\sigma(\phi_{S},\phi_\pi^H) &\sim & {\rm d}\sigma_0
+{\rm d}\Delta\sigma_0\sin\phi_{S}+
{\rm d}\sigma_1\cos\phi_\pi^H+ {\rm d}\sigma_2\cos2\phi_\pi^H+
{\rm d}\Delta\sigma_{1}^{-}\sin(\phi_{S}-\phi_\pi^H)
\nonumber\\
&+& {\rm d}\Delta\sigma_{1}^{+}\sin(\phi_{S}+\phi_\pi^H)
+{\rm d}\Delta\sigma_{2}^{-}\sin(\phi_{S}-2\phi_\pi^H)+
{\rm d}\Delta\sigma_{2}^{+}\sin(\phi_{S}+2\phi_\pi^H)\,,
\label{d-sig-phi-SA}
\end{eqnarray}
where $\phi_\pi^H$ is the azimuthal
angle of the pion three-momentum around the jet axis, as measured in the fragmenting parton helicity frame \cite{D'Alesio:2010am}.
In close analogy with the SIDIS case, in order to single out 
the different contributions of interest, we can define appropriate azimuthal moments,
\begin{eqnarray}
A_N^{W(\phi_{S},\phi_\pi^H)}(\bm{p}_{\rm j},z,k_{\perp\pi})
&=&
2\,\frac{\int{\rm d}\phi_{S}{\rm d}\phi_\pi^H\,
W(\phi_{S},\phi_\pi^H)\,[{\rm d}\sigma(\phi_{S},\phi_\pi^H)-
{\rm d}\sigma(\phi_{S}+\pi,\phi_\pi^H)]}
{\int{\rm d}\phi_{S}{\rm d}\phi_\pi^H\,
[{\rm d}\sigma(\phi_{S},\phi_\pi^H)+
{\rm d}\sigma(\phi_{S}+\pi,\phi_\pi^H)]}\,,
\label{gen-mom}
\end{eqnarray}
where $W(\phi_{S},\phi_\pi^H)$ is some appropriate circular function of $\phi_{S}$ and $\phi_\pi^H$.

\section{Phenomenology}

In this section we present and discuss some phenomenological implications of our approach in kinematic
configurations accessible at RHIC by the STAR and PHENIX experiments.
We consider both central ($\eta_{\rm j}=0$) and forward
($\eta_{\rm j}=3.3$) (pseudo)rapidity configurations
at a c.m.~energy $\sqrt{s} =$ 200 GeV (different c.m.~energies, namely $\sqrt{s}=62.4$ and 500 GeV, are also studied in \cite{D'Alesio:2010am}). 

For numerical calculations all TMD distribution and
fragmentation functions are taken in the simplified form where
the functional dependence on
the parton light-cone momentum fraction and on transverse
motion are completely factorized, assuming a Gaussian-like flavour-independent
shape for the transverse momentum component.
Concerning the parameterizations of the transversity and quark 
Sivers distributions,
 and
of the Collins functions, we consider two sets:
SIDIS~1 \cite{Anselmino:2005ea, Anselmino:2007fs} and  
SIDIS~2 \cite{Anselmino:2008sga,Anselmino:2008jk}. 
Furthermore, for the usual collinear distributions, we  adopt the LO 
unpolarized set GRV98~\cite{Gluck:1998xa}. For fragmentation functions, we adopt two well-known LO sets
among those available in the literature, the set by Kretzer~\cite{Kretzer:2000yf} and the DSS one~\cite{deFlorian:2007aj}.
Our choice is dictated by the subsequent use of the two available 
parametrization sets for the Sivers and Collins functions
in our scheme, that have been extracted in the past years by adopting these sets of fragmentation functions.
Since the range of the jet transverse momentum
(the hard scale) covered is significant, we  take into account proper evolution
with scale.  On the other hand, the transverse momentum component of all TMD functions
is kept fixed with no evolution with scale.
In all cases considered, $\bm{k}_{\perp\pi}$ is integrated over and, since we are interested in leading particles inside the jet, we  present results obtained
integrating the light-cone momentum fraction of the observed hadron, $z$,
in the range $z\geq 0.3$.

We have considered first, for $\pi^+$ production only,
a scenario in which the effects of
all TMD functions are over-maximized. By this we mean that all TMD
functions are maximized in size by imposing natural positivity bounds. 
The transversity distribution has been fixed at the initial scale by saturating the Soffer bound and then we let it
evolve. Moreover, the relative signs of
all active partonic contributions are chosen so that they
sum up additively. In this way we set
an upper bound on the absolute value of any of the effects playing
a potential role in the azimuthal asymmetries.
Therefore, all effects that are negligible or even
marginal in this scenario may be directly discarded in subsequent
refined phenomenological analyses. See Ref.~\cite{D'Alesio:2010am} for a more detailed discussion.

As a second step in our study we consider, for both neutral and charged pions,
only the surviving effects, involving TMD functions for which parameterizations
are available from independent fits to other spin and azimuthal
asymmetries data in SIDIS, Drell-Yan (DY), and $e^+e^-$ processes.
\begin{figure}[t]
\begin{center}
 \includegraphics[angle=0,width=0.4\textwidth]{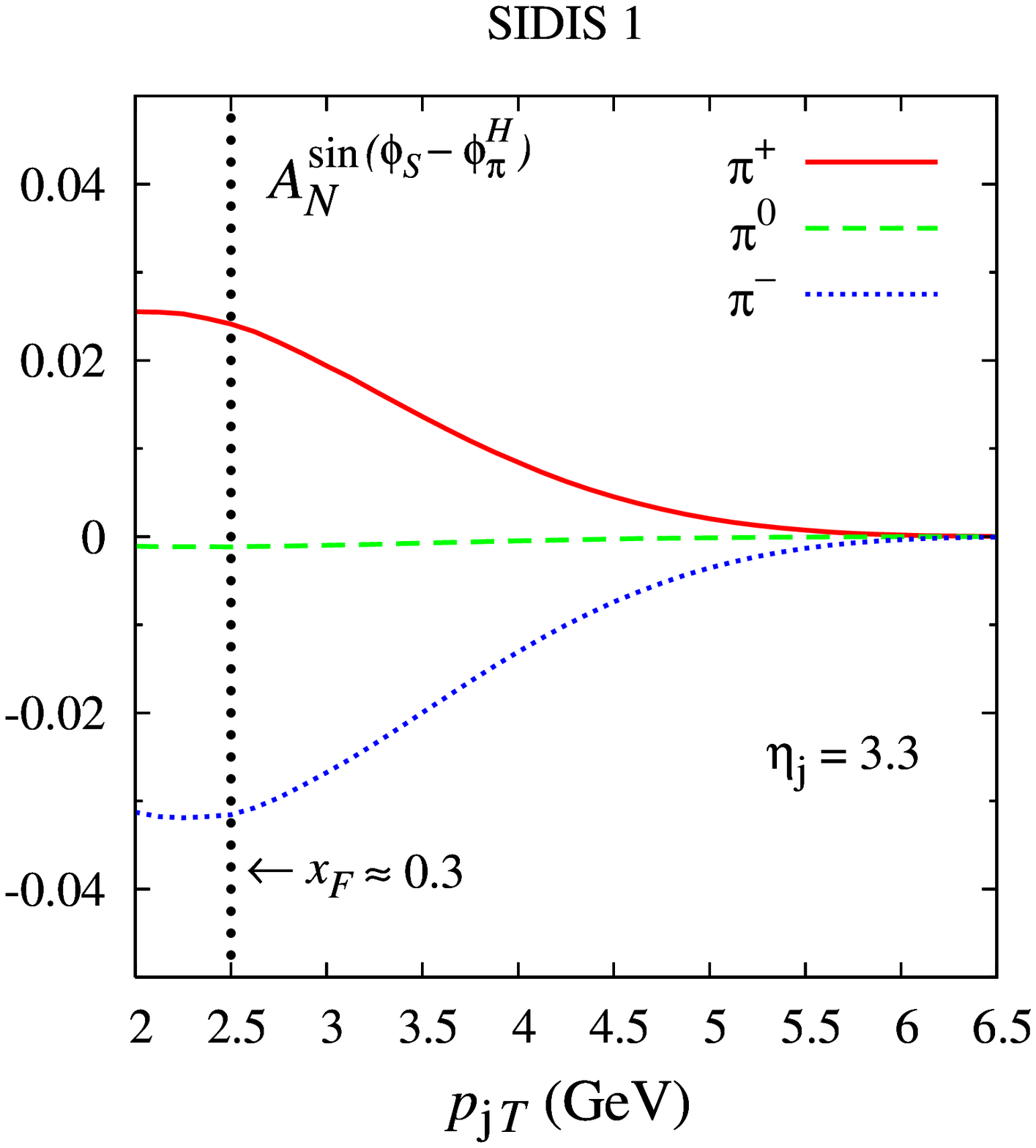}
 \includegraphics[angle=0,width=0.4\textwidth]{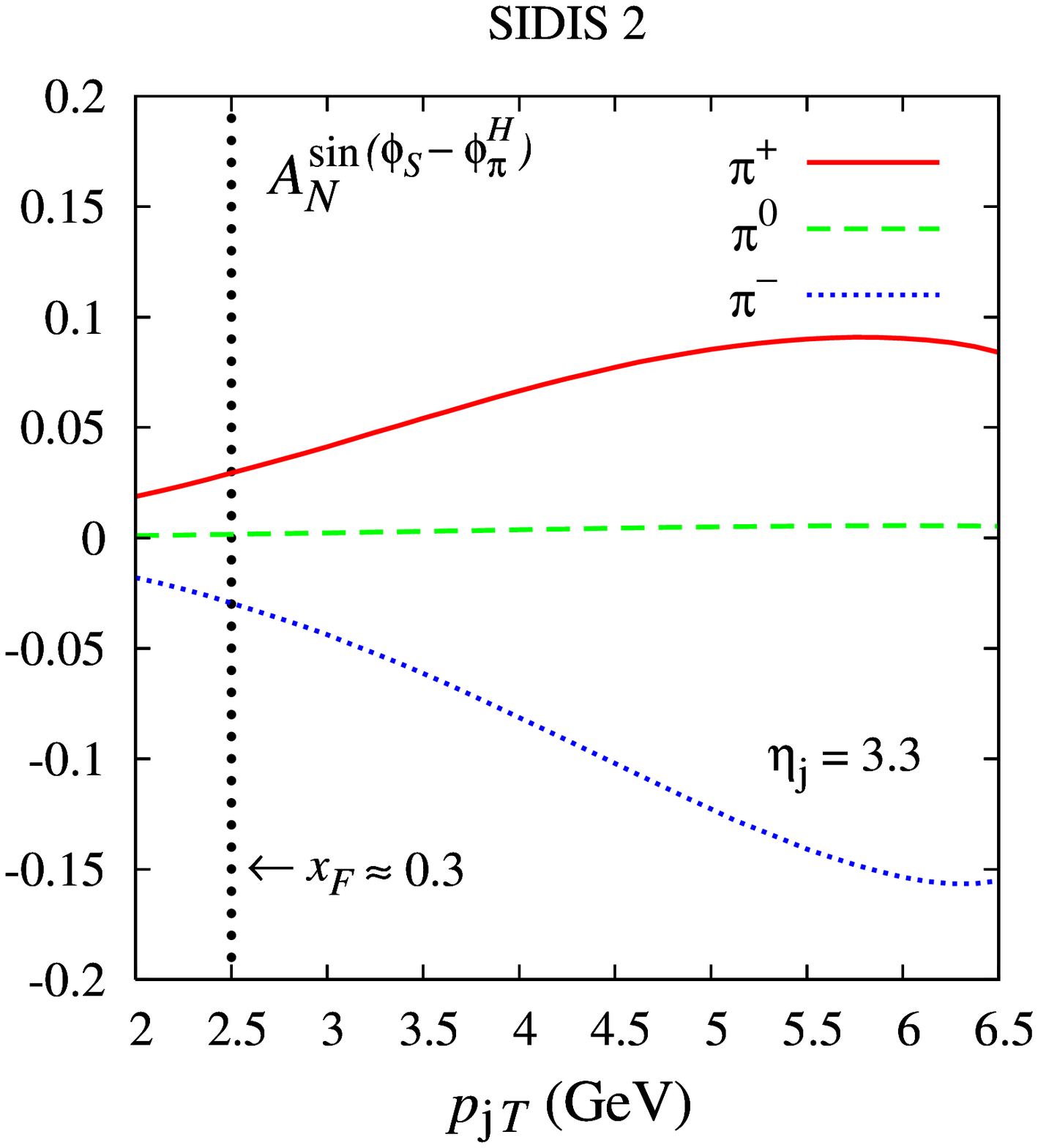}
 \caption{  The estimated quark Collins asymmetry
for the $p^\uparrow p\to {\rm jet}+\pi + X$ process,
obtained adopting the parameterizations SIDIS~1 
and SIDIS~2  respectively,
at $\sqrt{s}=200$ GeV  in the forward rapidity
region.
The dotted black vertical line delimits the region $x_F \approx 0.3$.
\label{asy-an-coll-par200} }
\end{center}
\end{figure}
\begin{figure}[b]
\begin{center}
 \includegraphics[angle=0,width=0.35\textwidth]{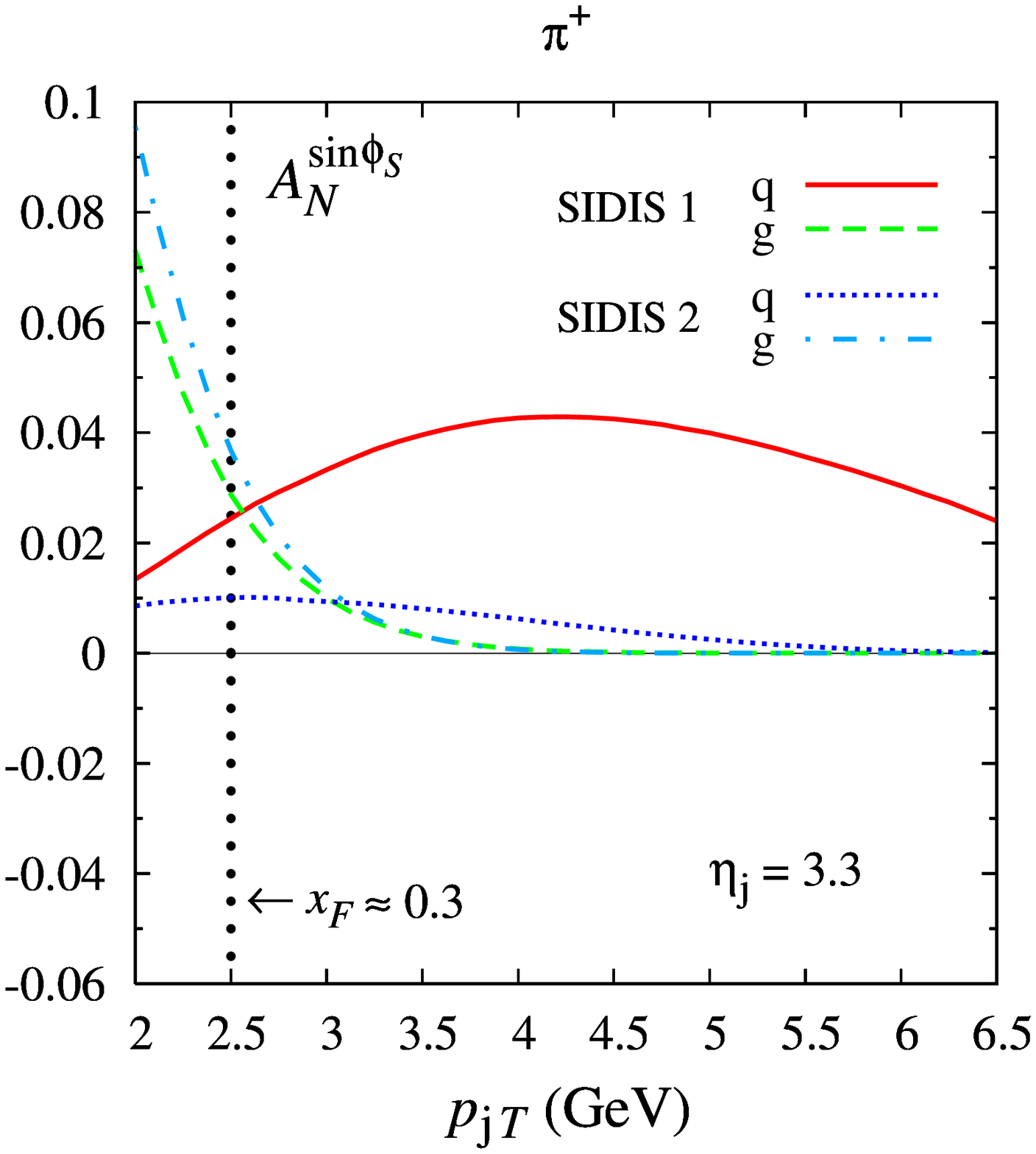}
 \hspace*{-20pt}
 \includegraphics[angle=0,width=0.35\textwidth]{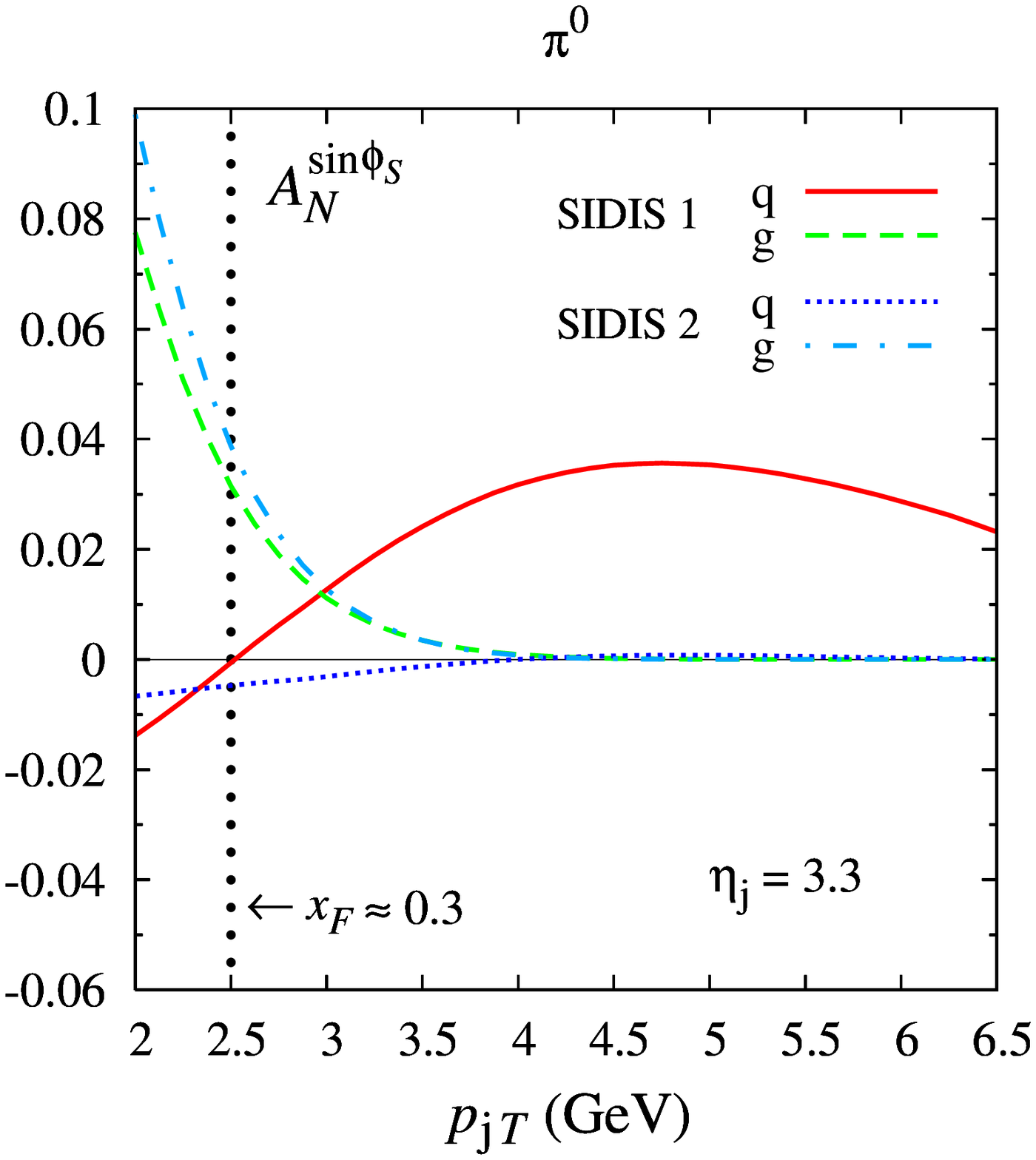}
 \hspace*{-20pt}
 \includegraphics[angle=0,width=0.35\textwidth]{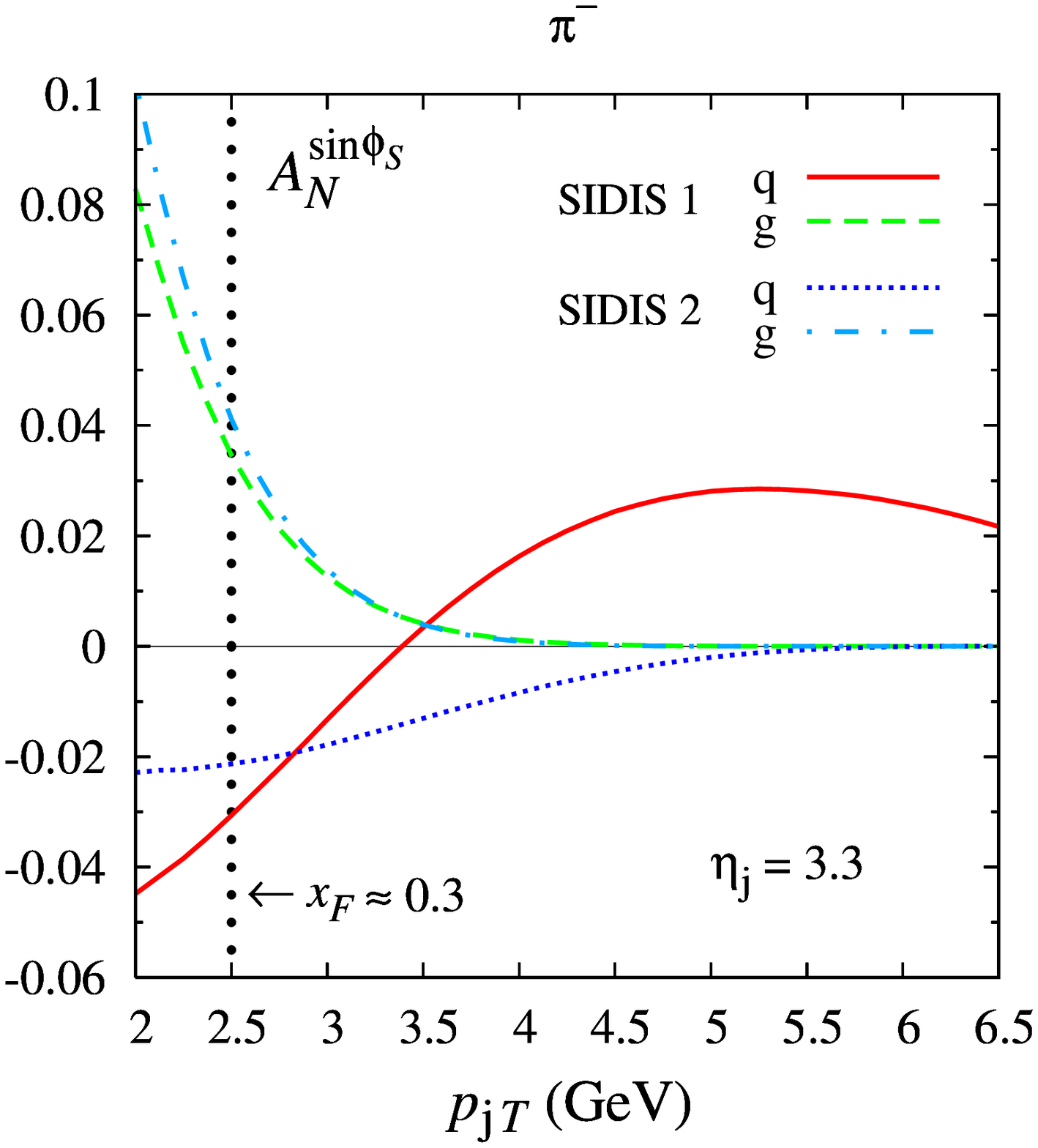}
 \caption{The estimated quark and gluon
 contributions to the Sivers asymmetry
 for the $p^\uparrow p\to {\rm jet}+\pi + X$ process,
 obtained adopting the parametrization sets
 SIDIS~1 and
 SIDIS~2,
at forward rapidity and $\sqrt{s}=200$ GeV.
The dotted black vertical line delimits the region $x_F\approx 0.3$.
 \label{asy-an-siv-par200} }
\end{center}
\end{figure}
In Fig.~\ref{asy-an-coll-par200}
we present, in the forward rapidity region, the quark generated asymmetry $A_N^{\sin(\phi_{S}-\phi_\pi^H)}$, which comes mainly from the convolution between the transversity distribution and the Collins fragmentation function. The plots are obtained  adopting the parameterizations SIDIS~1 and  SIDIS~2.
The Collins asymmetry for neutral pions
results to be almost vanishing, in agreement with preliminary RHIC data 
\cite{Poljak:2010tm,Poljak:2011vu}.
The dotted black vertical line delimits the region 
$x_F \approx 0.3$, with $x_F=2 p_{{\rm j}\,L}/\sqrt{s}$,
beyond which the SIDIS parameterizations for transversity are extrapolated 
outside the Bjorken $x$
region covered by SIDIS data and are therefore plagued by large uncertainties. For this reason, the two
parameterizations  for charged pions give comparable results
only below this limit (notice the difference in scale between the two panels).  A measurement of such asymmetries
would be then very important and helpful in clarifying
the large $x$ behaviour of the quark transversity distribution. Furthermore, it turns out that in the central rapidity region the quark Collins
asymmetries are practically negligible.

In Fig.~\ref{asy-an-siv-par200} we show,
for both neutral and charged pions, the quark and gluon
contributions to the Sivers asymmetry $A_N^{\sin\phi_{S}}$,
which cannot be disentangled, in the forward rapidity region as a function of $p_{{\rm j}\,T}$. The quark term is obtained adopting the SIDIS~1 and SIDIS~2 parameterizations. The almost unknown gluon Sivers function is tentatively taken positive and saturated to an updated version of the bound obtained in Ref.~\cite{Anselmino:2006yq} by considering PHENIX data for the $\pi^0$ transverse single spin asymmetry at mid-rapidity.
Similarly to the case of the Collins asymmetry, the two parameterizations
give comparable results only in the $p_{{\rm j}\,T}$ region where
 the behaviour of the quark Sivers distribution is reasonable well constrained by SIDIS data (see again the dotted black vertical line). 

\section{Test of the process dependence of the Sivers function}

In the GPM approach one applies the TMD distribution and fragmentation functions extracted from SIDIS, considering them to be universal. Very recently \cite{D'Alesio:2011mc} the azimuthal distribution of leading pions inside jets has been studied allowing for process dependence of the (quark) Sivers function. This is referred to as the colour gauge invariant (CGI) GPM  \cite{Gamberg:2010tj}. Namely, we have taken into account the effects of initial (ISIs) and final state interactions (FSIs) between the active parton and spectator remnants in the different scattering sub-processes. 
Since the details of such interactions depend on the particular partonic reaction considered, they render the Sivers function non-universal (see \cite{Gamberg:2010tj} and references therein). The oft-discussed case is the difference between the FSIs in SIDIS and the ISIs in DY scattering which leads to the prediction of an opposite relative sign of the Sivers functions in the two processes. This is considered to be a crucial test of our understanding of single spin asymmetries in QCD and still has to be confirmed by experiments.

\begin{figure}[t]
\includegraphics[angle=0,width=0.35\textwidth]{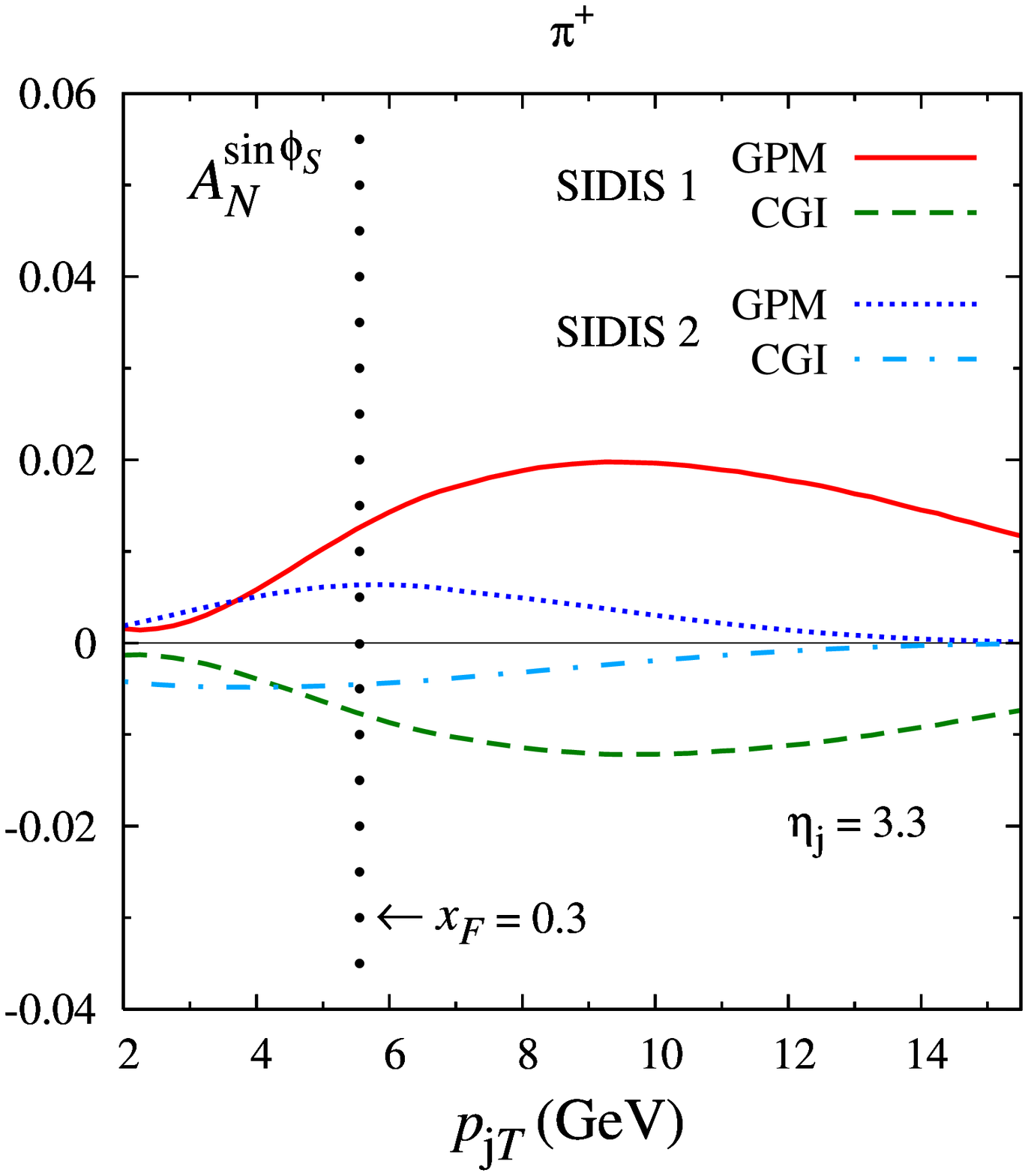}
 \hspace*{-20pt}
 \includegraphics[angle=0,width=0.35\textwidth]{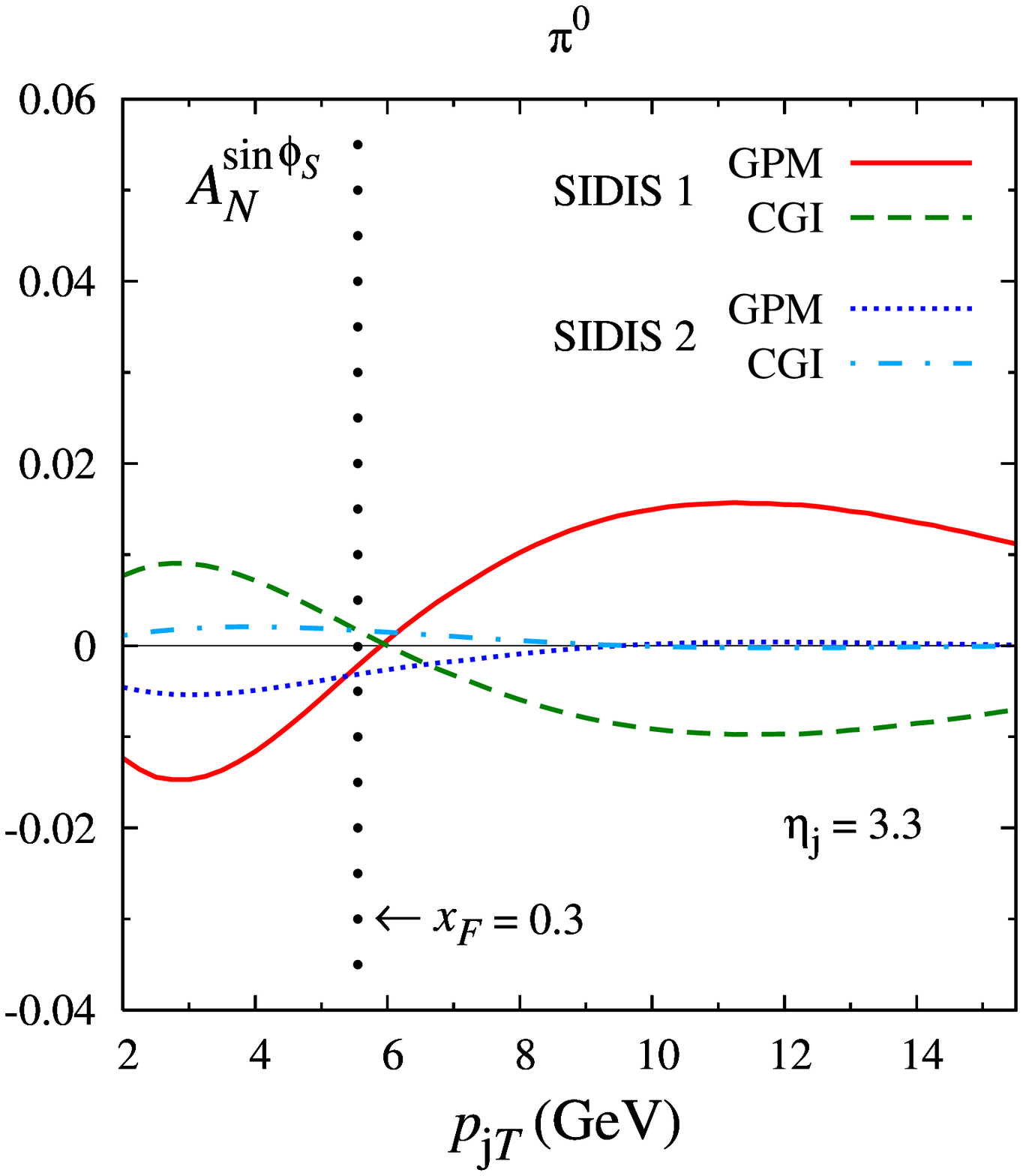}
 \hspace*{-20pt}
 \includegraphics[angle=0,width=0.35\textwidth]{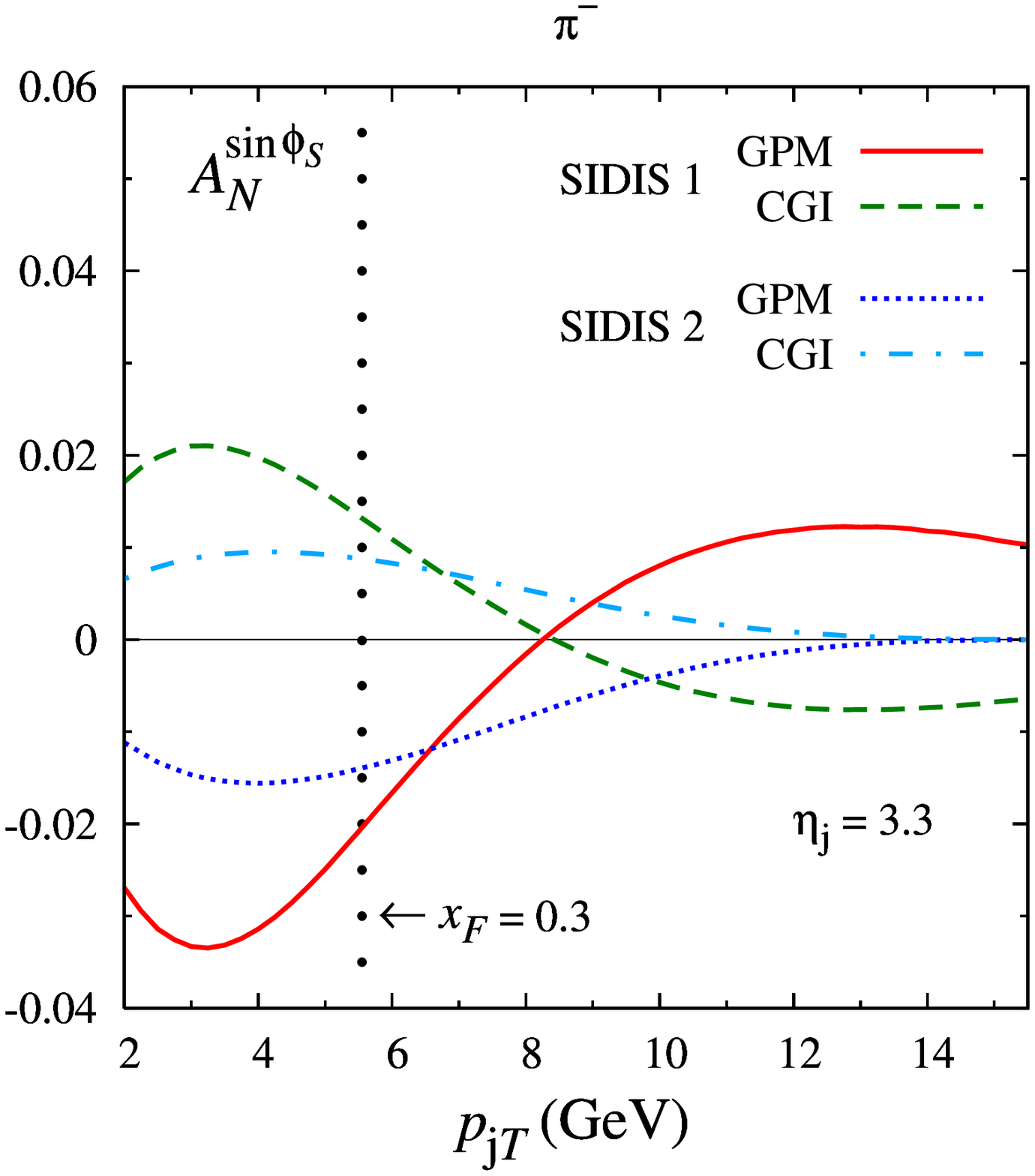}
\caption{The Sivers asymmetry $A_N^{\sin\phi_{S}}$ for the $p^\uparrow p\to {\rm jet} +\pi +X$ process
as a function of $p_{{\rm j}T}$, at fixed jet rapidity $\eta_{\rm j}=3.3$, for the RHIC energy $\sqrt{s}=500$ GeV.}
\label{fig1}
\end{figure}

 When applying similar reasoning to hadron production in $p\,p$ collisions, typically the Sivers function has a more complicated colour factor structure since both ISIs and FSIs contribute. However, in the forward rapidity region the process under study is dominated by only one channel, $qg\to qg$, with the final quark identified with the observed jet. As a consequence, one finds that the predictions for the Sivers asymmetries obtained with and without inclusion of color gauge factors are comparable in size but with {\em opposite signs} \cite{D'Alesio:2011mc}, similarly to the DY case.  Therefore the experimental observation of a sizable asymmetry could easily discriminate among the two approaches and test the universality properties of the Sivers function in hadronic reactions. 
Our results are shown in Fig.~\ref{fig1}, where we plot $A_N^{\sin\phi_
{S}}$ as a function of the jet transverse momentum $p_{{\rm j}T}$ at $\eta_{\rm j}=3.3$, for  the RHIC energy $\sqrt{s}=500$ GeV, integrated over $\bm{k}_{\perp\pi}$ and $z$ ($z\ge 0.3$).
The predictions labeled SIDIS~1 and SIDIS~2 are similar in the intermediate $p_{{\rm j}T}\le 5.5$ GeV region (corresponding to $x_F<0.3$). This region is then optimal to test directly the process dependence of the Sivers function.
Note that for $\sqrt s = 200$ GeV the behavior of our estimates would be similar to that shown in Fig.~\ref{fig1}, gaining almost a factor of 2 in size. However the range of $p_{{\rm j}T}$ covered would be narrower ($p_{{\rm j}T}\le 6.5$~GeV) and with $x_F \le 0.3$ now corresponding to $p_{{\rm j}T}\le 2.5$~GeV.

Finally, as a natural extension of this analysis, one can consider single inclusive jet asymmetry in $p\,p$ scattering, which is described solely by the Sivers function. The results obtained for $A_N^{\sin\phi_{S}}$ (not shown) look almost indistinguishable from the case of neutral pion-jet production (central panel of Fig.~\ref{fig1}).


\section{\label{sec-conclusions} Conclusions}
We have presented a study of the azimuthal asymmetries measurable
in the distribution of leading pions inside a large-$p_T$
jet produced in  single-transverse polarized proton proton
collisions for kinematic configurations accessible at RHIC.
To this end, we have adopted a generalized TMD parton model approach
with inclusion of spin and intrinsic parton motion effects both
in the distribution and in the fragmentation sectors. In contrast to inclusive  pion production, where the Sivers and Collins mechanisms cannot be separated,
 the leading-twist azimuthal asymmetries
discussed above would allow one to single out the different effects by taking suitable moments of the asymmetries. This will give us the opportunity of testing the
factorization and universality assumptions, and of gaining
information on the size and \textit{sign} of the various TMD functions in a kinematic region not covered by SIDIS data.

\acknowledgments
C.P.~is supported by Regione Autonoma della Sardegna under grant PO Sardegna FSE 2007-2013, L.R. 7/2007. U.D.~and F.M.~acknowledge partial support by Italian MIUR under PRIN 2008, and by the European Community (FP7 grant agreement No.~227431).


\begin{thebibliography}{20}

\bibitem{D'Alesio:2007jt}
\BY{D'Alesio~U. \atque Murgia~F.}
\IN{Prog. Part. Nucl. Phys.}{61}{2008}{394}.


\bibitem{Sivers:1989cc}
\BY{Sivers~D.W.}
\IN{Phys. Rev. D}{41}{1990}{83}; \IN{ibidem}{43}{1991}{261}.


\bibitem{Boer:1997nt}
\BY{Boer D. \atque  Mulders P.J.}
\IN{Phys. Rev. D}{57}{1998}{5780}.

\bibitem{Collins:1992kk}
\BY{Collins~J.C.}
\IN{Nucl. Phys. B}{396}{1993}{161}.

\bibitem{Anselmino:2005sh}
\BY{Anselmino~M. {\em et~al.}}
\IN{Phys. Rev. D}{73}{2006}{014020}.

\bibitem{D'Alesio:2010am}
\BY{D'Alesio~U., Murgia~F. \atque Pisano~C.} 
\IN{Phys. Rev. D}{83}{2011}{034021}.

\bibitem{Yuan:2007nd}
\BY{Yuan~F.}
\IN{Phys. Rev. Lett.}{100}{2008}{032003}.

\bibitem{Poljak:2010tm}
  \BY{Poljak~N. [STAR Collaboration]}
  \IN{J.\ Phys.\ Conf.\ Ser.}{295}{2011}{012102}.

\bibitem{Poljak:2011vu}
\BY{Poljak~N.~[STAR Collaboration]}
these proceedings, arXiv:1111.0755 (2011).


\bibitem{Anselmino:2005ea}
\BY{Anselmino~M.~{\em et~al.}}
\IN{Phys. Rev. D}{72}{2005}{094007}.

\bibitem{Anselmino:2007fs}
\BY{Anselmino~M.~{\em et~al.}}
\IN{Phys. Rev. D}{\bf 75}{2007}{054032}.

\bibitem{Anselmino:2008sga}
\BY{Anselmino~M.~{\em et~al.}}
\IN{Eur. Phys. J. A}{\bf 39}{2009}{89}.

\bibitem{Anselmino:2008jk}
\BY{Anselmino~M.~{\em et~al.}}
\IN{Nucl. Phys. Proc. Suppl.}{191}{2009}{98}.



\bibitem{Gluck:1998xa}
\BY{Gl{\"u}ck~M., Reya~E.  \atque Vogt~A.}
\IN{Eur. Phys. J. C}{5}{1998}{461}.

\bibitem{Kretzer:2000yf}
\BY{Kretzer~S.}
\IN{Phys. Rev. D}{62}{2000}{054001}.

\bibitem{deFlorian:2007aj}
\BY{de~Florian~D., Sassot~R., \atque Stratmann~M.} 
\IN{Phys. Rev. D}{75}{2007}{114010}.


\bibitem{Anselmino:2006yq}
\BY{Anselmino~M., D'Alesio~U., Melis~S. \atque Murgia~F.}
\IN{Phys. Rev. D}{74}{2006}{094011}.


\bibitem{D'Alesio:2011mc}
\BY{D'Alesio U.~{\em et~al.}}
\IN{Phys. Lett. B}{704}{2011}{637}.


\bibitem{Gamberg:2010tj}
\BY{Gamberg~L. \atque Kang~Z.B.}
\IN{Phys. Lett. B }{696}{2011}{109}.


\end{thebibliography}
\end{document}